\documentclass[journal,12pt,draftclsnofoot,onecolumn]{IEEEtran}
\ifCLASSINFOpdf

\else

\fi
\usepackage{cite}
\usepackage{mathrsfs}
\usepackage[english]{babel}
\usepackage{algorithm}
\usepackage{algorithmic}
\usepackage{amsmath}
\usepackage{amssymb}
\usepackage{hyperref}
\usepackage{nomencl}
\usepackage{subfigure,xcolor}

\usepackage{graphicx}
\usepackage[top=0.8in, margin=0.75in]{geometry}

\begin{document}
\begin{titlepage}
\begin{center}
\vspace*{-2\baselineskip}
\begin{minipage}[l]{7cm}
\flushleft
\includegraphics[width=2 in]{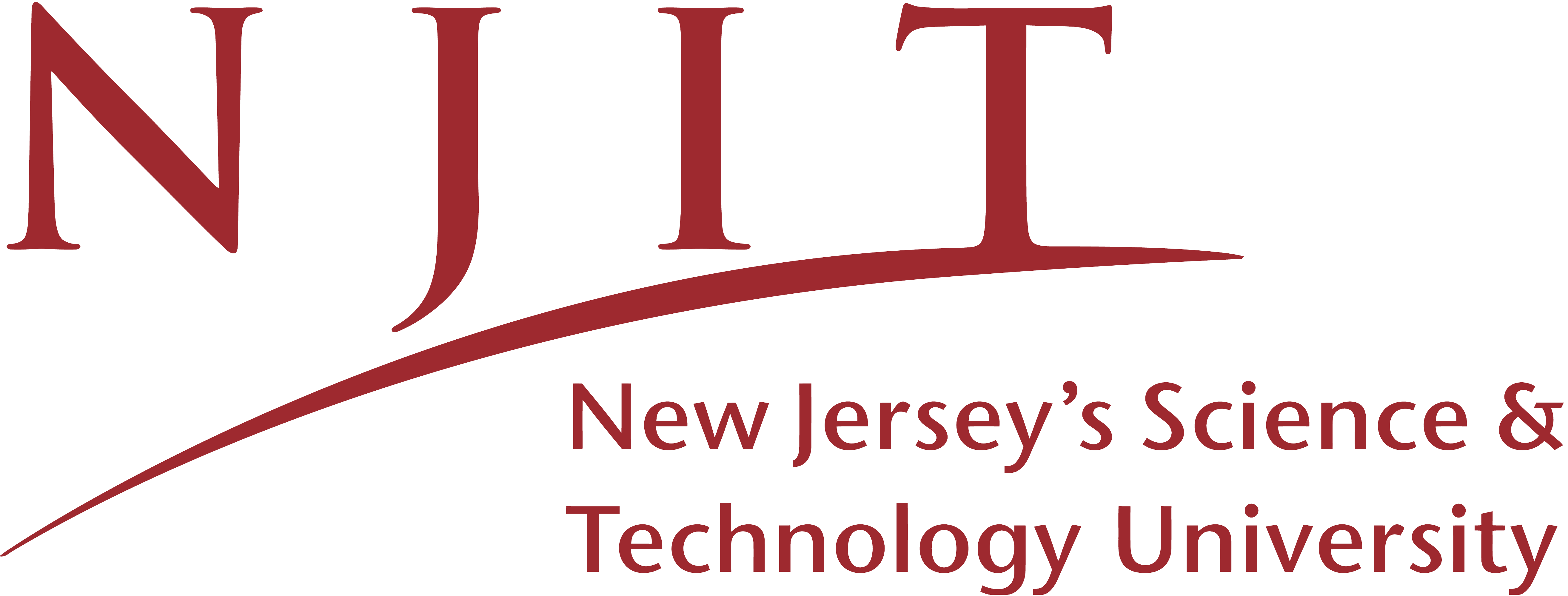}
\end{minipage}
\hfill
\begin{minipage}[r]{7cm}
\flushright
\includegraphics[width=1 in]{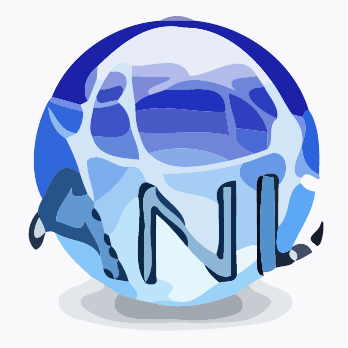}
\end{minipage}
\vfill
\textsc{\LARGE Learning-Assisted Secure End-to-End Network \\[12pt]
Slicing for Cyber-Physical Systems}\\
\vfill
\textsc{\LARGE QIANG LIU\\[12pt]
\LARGE TAO HAN \\[12pt]
\LARGE NIRWAN ANSARI}\\
\vfill
\textsc{\LARGE TR-ANL-2019-002\\[12pt]
\LARGE September 17, 2019}\\[1.5cm]
\vfill
{ADVANCED NETWORKING LABORATORY\\
 DEPARTMENT OF ELECTRICAL AND COMPUTER ENGINEERING\\
 NEW JERSY INSTITUTE OF TECHNOLOGY}
\vfill
\end{center}

\begin{minipage}[c]{16cm}
\flushleft
\small
{{Citation:}\\
Q. Liu, T. Han, and N. Ansari, ``Learning-Assisted Secure End-to-End Network Slicing for Cyber-Physical Systems," accepted for publication in the Special Issue, Cyber Security Based on Artificial Intelligence for Cyber-Physical Systems, in \emph{IEEE Network} to appear in May 2020.\\}
\end{minipage}

\end{titlepage}

\title{Learning-Assisted Secure End-to-End Network Slicing for Cyber-Physical Systems}
%
%
%

\author{Qiang~Liu,~\IEEEmembership{Student Member,~IEEE,}
        Tao~Han,~\IEEEmembership{Member,~IEEE,}
        and~Nirwan~Ansari,~\IEEEmembership{Fellow,~IEEE}
\thanks{Qiang Liu and Tao Han are with the Department
of Electrical and Computer Engineering, The University of North Carolina at Charlotte, NC,
28223, USA. E-mail: \{qliu12,Tao.Han\}@uncc.edu.}
\thanks{Nirwan Ansari is with Helen and John C. Hartmann Department of Electrical and Computer Engineering, New Jersey Institute of Technology, Newark, NJ, 07102, USA. E-mail: Nirwan.Ansari@njit.edu}
\thanks{ This work is partially supported by the U.S. National Science Foundation under Grant No. 1731675, No. 1810174, and No.1910844}}

\maketitle

\begin{abstract}
There is a pressing need to interconnect physical systems such as power grid and vehicles for efficient management and safe operations. Owing to the diverse features of physical systems, there is hardly a one-size-fits-all networking solution for developing cyber-physical systems. Network slicing is a promising technology that allows network operators to create multiple virtual networks on top of a shared network infrastructure. These virtual networks can be tailored to meet the requirements of different cyber-physical systems. However, it is challenging to design secure network slicing solutions that can efficiently create end-to-end network slices for diverse cyber-physical systems. In this article, we discuss the challenges and security issues of network slicing, study learning-assisted network slicing solutions, and analyze their performance under the denial-of-service attack. We also present a design and implementation of a small-scale testbed for evaluating the network slicing solutions.
\end{abstract}

\IEEEpeerreviewmaketitle

\section{Introduction}
\IEEEPARstart{A}{n} essential feature of cyber-physical systems is to connect physical devices and infrastructure such as autonomous vehicles and micro power grid to the Internet for efficient system control, management and monitoring~\cite{Yu:2016:CPS}. Since different physical systems have diverse requirements of network resources, there is hardly a one-size-fits-all networking solution for cyber-physical systems. It is also impractical to deploy customized network infrastructure and protocols for each cyber-physical system. Therefore, how to efficiently connect heterogeneous physical systems to the Internet in a cost-effective way is still an open problem.

Networking slicing emerges as a promising technology for serving the specific needs of vertical industries~\cite{Katsalis:2017:NST}. The network slicing technology empowers mobile network operators to create multiple virtual networks, i.e., network slices, on top of shared physical network infrastructure~\cite{5GSlicing}. The virtual network can be customized to satisfy a variety of requirements of network performance and functionality. For instance, a network slice can be created to support smart grid communications with ultra-low latency and high reliability. Meanwhile, since smart grid control usually does not need to transfer a large amount of data, the slice can be customized with low throughput.

To support compute-intensive applications such as machine learning and artificial intelligence,
an increasing number of cyber-physical systems require powerful computing infrastructure. For example, autonomous vehicles need high computation capability to analyze the data collected from various sensors such as LIDAR (Light Detection and Ranging) and cameras in a real-time fashion. Since the in-vehicle computation often radiates heat that can dramatically increase the temperature inside the car, it is desirable to offload the compute-intensive tasks to edge computing infrastructure~\cite{MEC5G,ansari2018mobile}. Hence, connecting modern physical systems usually needs resources from multiple technical domains such as radio access networks and computing servers.

The main difficulty in network slicing lies in how to utilize the physical network and computing infrastructure efficiently and provide reliable and secure connection and computation to cyber-physical systems. Many conceptual network slicing frameworks have been proposed by researchers from both academia and industry~\cite{ksentini2017toward,rost2017network,Katsalis:2017:NST,zhang2017network,5GSlicing}. However, only a few papers provide the in-depth discussion of network slicing algorithms~\cite{kokku2012nvs,kokku2013cellslice,Caballero:2017:NSG} and present realizable system designs~\cite{foukas2016flexran,foukas2017orion}. Although these papers provide useful insights on network slicing and lay foundations for prototyping network slicing solutions, they solely focus on slicing radio access networks and do not consider the performance of a network slice which requires multiple resources, e.g., radio and computation resources. Moreover, none of these papers designs network slicing algorithms and systems with consideration of multiple radio access points and edge servers. In addition, existing works fail to evaluate the reliability and vulnerability of network slicing solutions.

In this article, we discuss the challenges of end-to-end network slicing that involves multi-domain resource orchestration for heterogeneous cyber-physical systems. Then, we study learning-assisted network slicing solutions~\cite{Liu:2019:vEdge,Liu:2019:DIRECT} and analyze their performance under the denial-of-service (DoS) attack. Finally, we present the software and hardware required for developing the network slicing testbed.

The remainder of the article is organized as follows. Section~\ref{sec:challenges} discusses the challenges of end-to-end network slicing for cyber-physical systems. Section~\ref{sec:las_net_slicing} presents learning-assisted end-to-end network slicing solutions. Section~\ref{sec:sim_results} evaluates the performance of the solution under the DoS attack through simulations. Section~\ref{sec:implementation} shows the design and implementation of the proposed solution on a small scale testbed. Section~\ref{sec:conclusion} discusses the future research directions and concludes the article.

\begin{figure}[!t]
	\centering
	\includegraphics[width=0.9\linewidth]{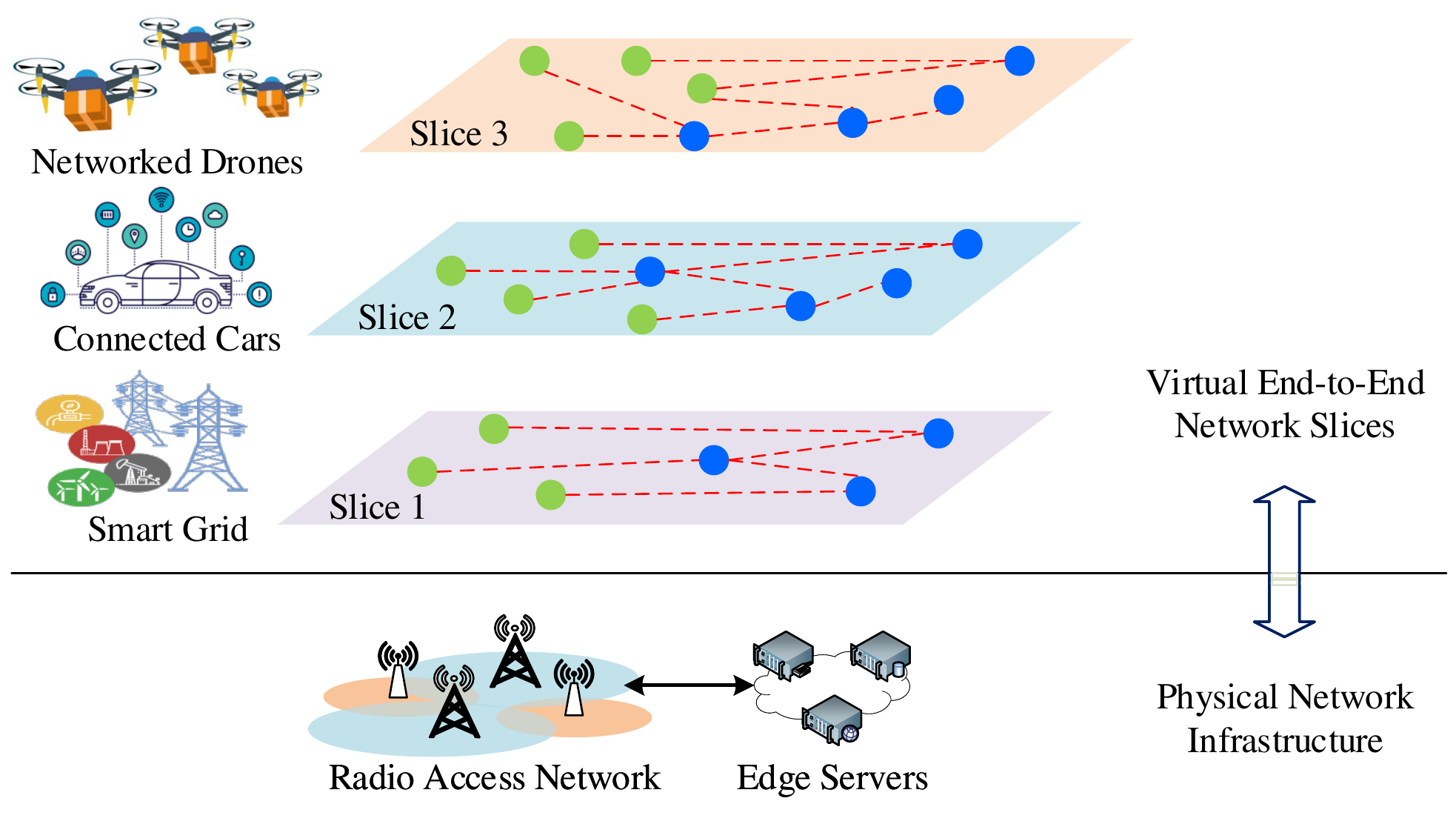}
	\caption{\small End-to-end network slicing for cyber-physical systems.}
	\label{fig:cps_slicing}
\end{figure}
\section{Challenges of End-to-End Network Slicing}
\label{sec:challenges}
In this section, we discuss the challenges of end-to-end network slices for cyber-physical systems. Fig.~\ref{fig:cps_slicing} provides an example of network slicing for three cyber-physical systems: smart grid, connected cars, and networked drones. Here, there are two parties: service providers and network operators. The service provider aims to create network slices to connect its physical systems, and the network operator owns and manages its network infrastructure. The service provider requests the network operator to create network slices and will, once instantiated, manage them. Given the requests from multiple service providers, the network operator instantiates network slices to meet the diverse requirements of service providers while optimizing the utilization of the network infrastructure.

\subsection{Heterogeneous resource demand v.s. slice performance}
Modern cyber-physical systems require a variety of cyber resources from multiple technical domains. For example, autonomous cars need communication and computation resources to transfer and analyze sensor data, respectively. The fundamental research challenge of slicing network resources for cyber-physical systems is from the difficulty in determining how the resource allocation in each technical domain impacts the performance of a network slice. Some cyber-physical systems, e.g., smart grid, require ultra-reliable and low-latency transmission but few computation resources. Some cyber-physical systems such as connected cars and networked drones need both low-latency communication connections and high-performance computation resources. Since cyber-physical systems have diverse requirements on different resources, the network operator is unable to develop a slice performance model that correctly characterizes the slice performance versus the resource allocation in different technical domains. As a result, it is challenging to orchestrate multi-domain resources to build a network slice for a cyber-physical system.

Cyber-physical systems are usually deployed over a large area, and require a collection of communication and computation infrastructure that can cover the area. That is to say, a network slice consists of many radio access points and edge servers. When creating a network slice, the network operator needs to consider the spatial diversity of the traffic loads generated from cyber-physical systems and allocate the resource properly among radio access points and edge servers to ensure the performance of cyber-physical systems and support seamless mobility. Unfortunately, the fact that the traffic and workloads of cyber-physical systems are time-variant further complicates the network slicing problem.

\subsection{Isolation v.s. utilization}
In general, there are two objectives in network slicing. The first one is to optimize the utilization of network and computation infrastructure in order to maximize the profit of network operators. The second one is to enforce the performance and functional isolation among network slices in order to ensure the performance of network slices. The performance isolation guarantees that the performance of a network slice will not affect or be affected by other network slices created on the same network and computation infrastructure, and the functional isolation allows service providers to customize their network slices and control their network operations independently~\cite{foukas2017orion}.

There is, however, a conflict between isolation and resource efficiency. In wireless communications, it is important to leverage diversity gains such as frequency diversity and multi-user diversity to improve the efficiency of radio resources and mitigate dynamic channel fading. Exploring the diversity gain requires pooling the resources together. The diversity gain fades away as the resources are sliced into pieces for isolation. Therefore, functional and performance isolation may reduce the efficiency of utilizing the resources.

The functional isolation provides the service providers, i.e., cyber-physical systems, the flexibility in managing their virtual network and computation resources. As a result, service providers can customize their slice operations such as traffic load balancing and user scheduling. The customized slice management strategies change the demands of communication and computation resources across networking and computing infrastructure. With the functional isolation, optimizing the network slicing requires the network operator to learn the customized management strategies and traffic profiles of individual network slices. Sharing the information about the management strategies and traffic profiles with network operators will incur excessive communication overhead and is not practical.

\subsection{Virtualization v.s. security}
Network slicing may introduce new vulnerabilities to cyber-physical systems. Network slicing enables network operators to manage networking and computing infrastructure, and service providers to control the operations of individual network slices. When creating a network slice, network operators allocate resources from multiple technical domains to serve a cyber-physical system. These resources are virtual and instantiated on physical networking and computing infrastructure. The service providers, i.e., cyber-physical systems, manage the virtual resources to maximize their utilities.

When an attacker launches an attack, e.g., denial-of-service (DoS), toward the network infrastructure, it is very difficult for network operators to detect the attack because they do not know how the service provider utilizes the resources and whether the traffic loads are legitimate or not. The service providers are also unable to detect the attack because they only manage the virtual resources and have no information about the mapping from virtual to physical resources. When the attack happens, the performance of affected network slices degrades. However, the service provider may recognize the attack as a change of mapping from virtual to physical resources, i.e., the inflation of virtual resources. As a result, service providers may request more virtual resources from the network operators.
The network operators may treat such requests as the traffic load increases in network slices rather than recognizing them as abnormal behaviors of the network slice.

\begin{figure*}[!t]
	\centering
	\includegraphics[width=5.0in]{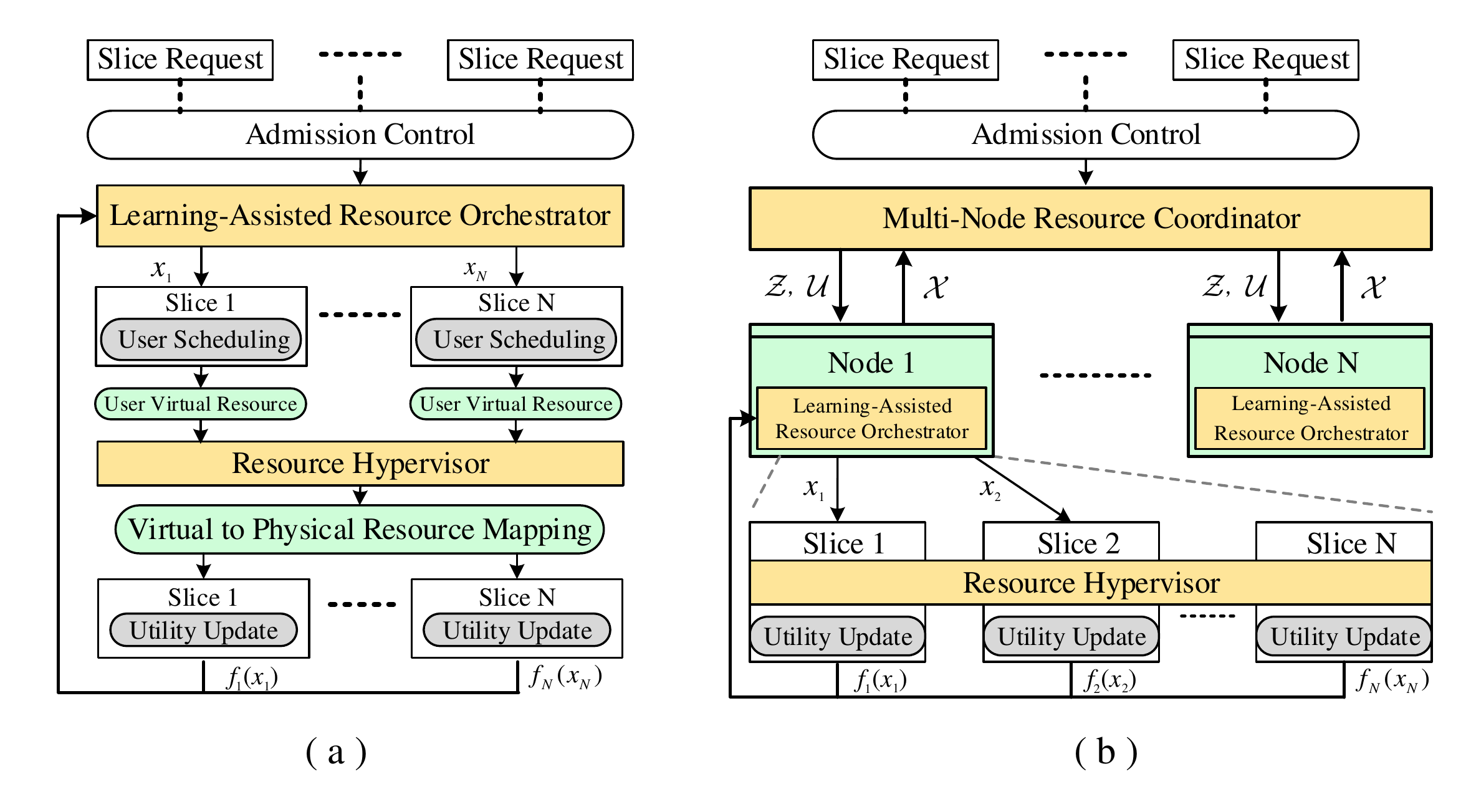}
	\vspace{-0.1in}
	\caption{\small The illustration of network slicing procedures for (a)  a single network node~\cite{Liu:2019:vEdge} and (b) multiple network nodes~\cite{Liu:2019:DIRECT}.}
	\label{fig:protocol}
\end{figure*}

\section{Learning-Assisted Secure Network Slicing}
\label{sec:las_net_slicing}
The security vulnerability of network slicing for cyber-physical systems is due to the lack of information sharing between the network operator and service provider. However, sharing the information, e.g., resource management strategies and traffic load profiles, is not practical because of the excessive communication and computation overhead. In this section, we discuss learning-assisted network slicing methods that allow the network operator to learn the performance of a network slice under given resource allocation. The learning results help the network operator to understand how the service providers, i.e., cyber-physical systems, utilize the communication and computation resources and what the utilities of the network slice are. The network operator may leverage such learning results to detect malicious attacks toward its network infrastructure and adjust its resource orchestration solutions to mitigate the impact of the attack on the performance of network slices. We first study the network slicing solution with consideration of a single network node and then extend the solution to create network slices over multiple network nodes. Here, we assume that a network node is composed of both networking and computation resources.

\subsection{Network slicing on a single network node}
The network slicing solution for a single network node is to efficiently utilize the networking and computation resources while ensuring the performance and functional isolation among network slices~\cite{Liu:2019:vEdge}. As shown in Fig.~\ref{fig:protocol} (a), the network slicing solution consists of two main components: learning assisted resource orchestrator and resource hypervisor.

\textbf{Learning-assisted resource orchestrator:} the resource orchestrator is responsible to orchestrate the resource allocation in multiple technical domains to support services in network slices. Owing to the diverse resource demands of cyber-physical systems, the resource orchestrator is unable to model the relationship between the slice performance and multi-domain resource allocation. Therefore, the orchestrator adopts a probabilistic model to represent the slice performance function, $f_{i}(x_i)$, of the $i$th slice under different resource allocation, $x_i$, and exploits the model to learn the properties of the function. Based on the learning results, the orchestrator estimates the gradient of the performance function for each slice and optimizes the resource allocation among the slices by using the proximal gradient method.

\textbf{Resource hypervisor:} the function of the resource hypervisor is to map the virtual resources to communication and computation resources in the network node. In the virtual-to-physical resource mapping, the resource hypervisor knows the channel state information of the users scheduled on the virtual resources. Therefore, the hypervisor can exploit the diversity gains in wireless communications to improve the efficiency of the radio resources.

\textbf{Network slicing procedure:} Fig.~\ref{fig:protocol} (a) illustrates the network slicing procedure on a single network node. The service providers send slice requests to the network operator to create network slices. Based on the available resources and service level agreement, the network operator admits selected slice requests. Then, the learning-assisted resource orchestrator allocates multi-domain resources to network slices to support their services. The resources allocated to network slices are virtual resources. The service providers can customize their resource management strategies and schedule traffic loads on the virtual resources. Afterward, the resource hypervisor maps the virtual resources to networking and computing infrastructure.

\textbf{Security analysis:} the learning-assisted resource orchestrator is able to detect a DoS attack by tracking the properties of the slice performance function. When a network slice experiences the DoS attack, given the same resource allocation, the performance of the slice will be degraded. By learning the properties of the slice performance function, the resource orchestrator will observe dramatic changes in the efficiency of the resource utilization in the slice, and thus detect the DoS attack. Then, the resource orchestrator will reduce the resource allocation to the slice and thus mitigate the impact of the attack.

\subsection{Network slicing over multiple network nodes}
With consideration of multiple network nodes, the network operator needs to properly allocate resources to each network node to meet the coverage requirements of cyber-physical systems and support mobility.
To this end, we design a new network slicing solution which integrates the alternating direction method of multipliers (ADMM) method, a learning-assisted optimization algorithm and the multi-domain resource hypervisor~\cite{Liu:2019:DIRECT}. In the solution, the network slicing problem is decomposed into subproblems that can be solved by individual network nodes based on the ADMM method. Since the total amount of resources can be allocated to a network slice is determined by the service level agreement, a multi-node resource coordinator is designed to coordinate resource orchestration among network nodes and enforce the service level agreement.

\textbf{Multi-node resource coordinator:} the coordinator controls the multi-domain resource orchestration in network nodes and enforces network slices to be served based on their service level agreement with the network operator. As shown in Fig.~\ref{fig:protocol} (b), the multi-node resource coordinator learns the performance of network slices on each network node via the resource allocation report, $\mathcal{X}$, and controls the resource orchestration by adapting the auxiliary variables, $\mathcal{Z}$, and the variables, $\mathcal{U}$. On each network node, the learning-assisted resource orchestrator incorporates $\mathcal{Z}$ and $\mathcal{U}$ in allocating resources to network slices.

\textbf{Security analysis:}
the multi-node resource coordinator helps mitigate the impact of malicious attacks toward a network node by controlling the resource allocation to the node. For example, if a network experiences the DoS attack,  the auxiliary variables, $\mathcal{Z}$, and the variables, $\mathcal{U}$, reported by the learning-assisted resource orchestrator will be changed. In general, such a change informs the multi-node resource coordinator that allocating resources to the network does not improve the performance of the network slices. As a result, the multi-node resource coordinator will reduce the resource allocation to the network node and re-balance the resource distribution among other network nodes that can meet the requirements of the network slices. Eventually, no network slice subjected to the DoS attack will be hosted on the network node.

\begin{figure*}[!t]
	\centering
	\includegraphics[width=5.0in]{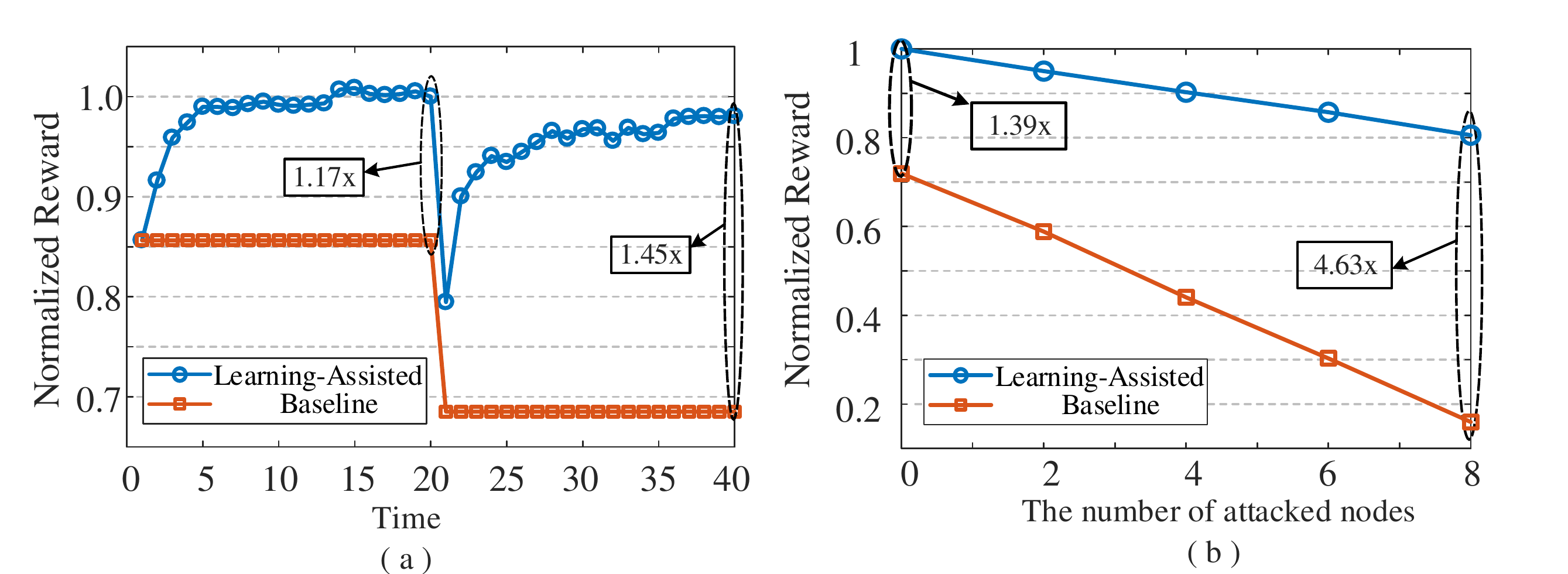}
	\vspace{-0.1in}
	\caption{\small The simulation results: (a) the performance versus time, and (b) the performance versus the number of attacked nodes.}
	\label{fig:simulation_results}
\end{figure*}

\section{Slice Performance under DoS Attack}
\label{sec:sim_results}
In this section, we perform network simulations to evaluate the performance of the learning-assisted network slicing solution under the DoS attack.
In the simulation, there are 5 network nodes, and each node consists of 5 users.
For supporting cyber-physical systems, a network slice is composed of three types of resources: uplink and downlink radio, and computation resources. The total amount of each resource is 100 units. We assume that the utility function of the $i$th slice in the $j$th network node is
$r_{i,j} = \sum \nolimits_{k \in \mathcal{K}}  \alpha_k \cdot (x_{i,j,k})$
where $x_{i,j,k}$ is the $k$th resource of the $i$th slice in the $j$th network node. $\alpha_k$ is the weight for the $k$th resource and generated according to a uniform distribution ranging from 1 to 10. We compare the performance of the learning-assisted algorithm with a baseline algorithm which allocates all resources evenly among all the network slices and distributes the resources of a network slice evenly to all network nodes.

Fig.~\ref{fig:simulation_results} (a) shows the performance of the learning-assisted algorithm under the DoS attack. The attack is launched toward one node at the $20$th time slot.
In the beginning, the learning-assisted algorithm appropriates the same resource allocation as the baseline algorithm does.
Then, the learning-assisted algorithm gradually learns the slice performance functions and improves the overall utilities by optimizing the resource allocation among nodes and slices.
The learning-assisted algorithm converges after the $6$th time slot time and obtains 1.17x performance improvement as compared to the baseline algorithm.
Once the attack on a node occurs, the performance of network slices significantly decreases under both the learning-assisted algorithm and baseline algorithm.
The algorithm is able to learn the changes of the resource utilization efficiency on each node with respect to the slice performance.
The learned results help to detect the attack on nodes and further adjust the resource allocation among nodes.
For example, the algorithm allocates more resources toward the nodes with higher resource utilization efficiency and decrease the resource provision of nodes with lower resource utilization efficiency.
In this way, the malicious attack on the node can be excluded from the network.
Since the resources are favorably allocated to high efficiency nodes, the learning-assisted algorithm mitigates the impact of the DoS attack and restores nearly 98\% of the performance of the network slices.
In addition, under the DOS attack, the slice performance with the learning-assisted algorithm is 1.45x better than that with the baseline algorithm.

Fig.~\ref{fig:simulation_results} (b) shows the performance of the network slices when the number of network nodes instigated by the DoS attack increases. The total number of network nodes in the simulation is 10.
Without attacks, i.e., the learning-assisted algorithm obtains 1.39x better performance than the baseline algorithm. When the number of the network nodes experiencing the DoS attack increases, the performance of the network slices decreases under both algorithms. However, the learning-assisted algorithm is able to minimize the impact of the attack on the performance of the network slices. For example, when 8 network nodes are attacked, the learning-assisted algorithm can identify the under-attack nodes and adjust the resource allocation among nodes to exclude the malicious attacks in the network. As a result, the slice performance obtained by the learning-assisted algorithm is 4.63x better than that with the baseline algorithm.

These simulation results validate the learning-assisted network slicing solution to be able to mitigate the impact of the DoS attack on the performance of the network slices. In other words, the learning-assisted network slicing solution can create network slices that are reliable and secure for cyber-physical systems.

\begin{figure*}[!t]
	\centering
	\includegraphics[width=7.0in]{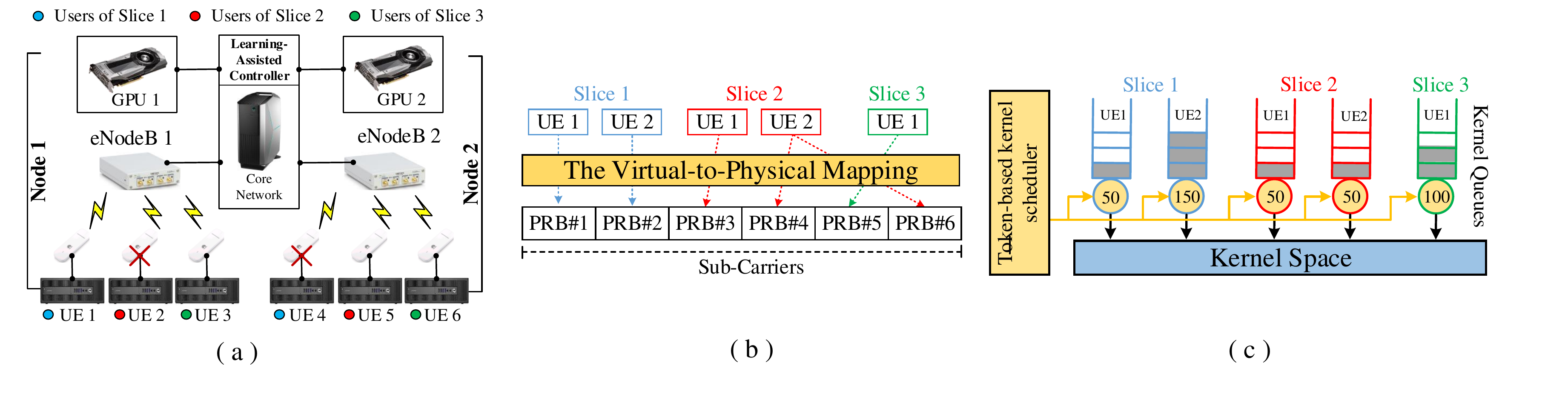}
	\vspace{-0.2in}
	\caption{\small The system design: (a) The system prototype, (b) radio resource hypervisor, and (c) computing resource hypervisor~\cite{Liu:2019:DIRECT}.}
	\label{fig:implementation}
	\vspace{-0.1in}
\end{figure*}

\section{System Prototyping and Results}
\label{sec:implementation}
In this section, we present the design of a small-scale prototype for evaluating the end-to-end network slicing solutions.
\subsection{Prototype Design}
\textbf{System Hardware:}
in the prototype, we consider the radio communication network and GPU computing platform as the main components.
As shown in Fig.~\ref{fig:implementation}, the prototype consists of two network nodes, and each node has both radio and computing resources.
The radio access network and core network are implemented based on the OpenAirInterface (OAI) LTE platform and openair-cn\footnote{OpenAirInferace is an open-source platform and implementation of 3GPP cellular networks. Available online: \url{https:gitlab.eurecom.fr/oai}}, respectively.
We deploy two eNodeBs in different places to emulate a cellular network with limited co-channel interference.
The computing platform is built based on NVIDIA CUDA-enable GPU\footnote{CUDA is a GPU parallel computing architecture developed by NVIDIA.}.
We use a computer with two NVIDIA GTX 1080Ti as the computing platform.
Ettus USRP B210 SDR is adopted as the RF front-end of an eNodeB, and LTE dongles are used to emulate mobile users.

\textbf{Radio Resource Hypervisor:}
the radio resource hypervisor maps the virtual radio resources to physical radio resources in LTE networks, i.e., physical resource blocks (PRBs) of PUSCH/PDSCH.
Here, we define the virtual resource as radio bandwidth that can be flexibly allocated to users by network slices, e.g., 360kHz.
We let network slices on a node share the same control plane following the LTE standards, and focus on allocating the uplink/downlink PRB resources in the user plane.
As illustrated in Fig.~\ref{fig:implementation} (b), the radio resource hypervisor maps users' virtual radio resources to PRBs.
Since the user information, i.e., channel condition and virtual resources, is known during the mapping, we leverage the information to maximize the network throughput.
In particular, we greedily select the user with the best channel condition for each PRB.

\textbf{Computing Resource Hypervisor:}
the computing resource hypervisor maps virtual computing resources to the GPU computing resources. In the prototype, we use the CUDA programming model, in which an application can invoke multiple \textit{kernels}, and executing each \textit{kernel} requires a number of CUDA threads. To manage the computing resource, we develop a token-based kernel scheduler to control the execution of \textit{kernels}. Here, the number of tokens reflect the amount of virtual computing resources. That is, a user with more tokens is able to use more computing resources. As illustrated in Fig.~\ref{fig:implementation} (c), the kernel scheduler dispatches the \textit{kernels} according to the available tokens of users. We develop a \textit{KernelSpawn} function to manage users' \textit{kernels} as a FIFO queue. Once a user has sufficient tokens, the user's \textit{kernel} is pulled out of the queue and executed.

\begin{figure}[!t]
\centering
	\includegraphics[width=0.9\linewidth]{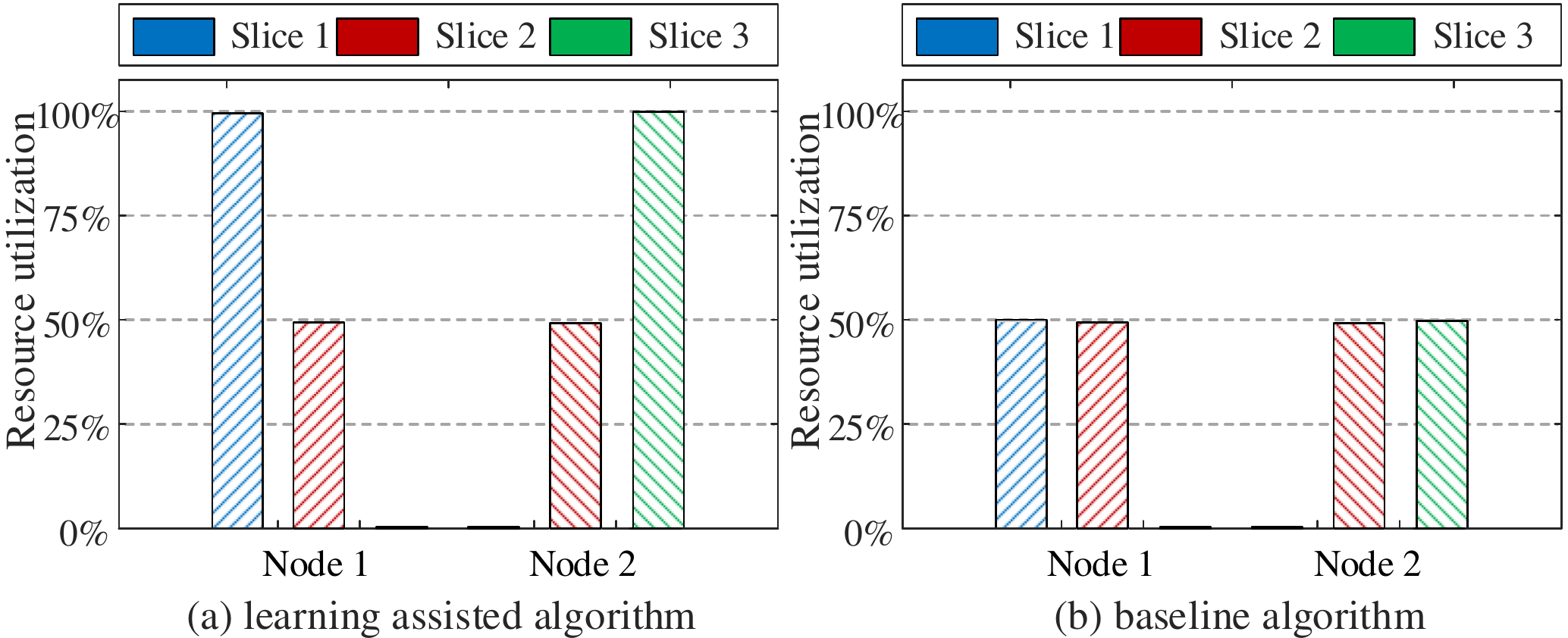}
	\caption{\small The performance of network slicing solutions under the DoS attack.}
	\label{fig:sys_allocation_detail}
	\vspace{-0.1in}
\end{figure}

\subsection{Experimental Results}
With the system prototype, we evaluate the performance of the learning-assisted algorithm under the DoS attack. In the experiment, we create three network slices over two network nodes to serve six mobile users. Each network node hosts three network slices, and one user is associated with a network slice on a network node. In the experiment, the DoS attack is launched toward network slice 1 and 3 on network nodes 2 and 1, respectively. (see Fig.~\ref{fig:implementation}.)

Fig.~\ref{fig:sys_allocation_detail} shows resource allocated to network slices on different nodes with the baseline and learning-assisted algorithms.
Fig.~\ref{fig:sys_allocation_detail} (a) shows that all resources of network slices 1 and 3 are allocated by the learning-assisted algorithm to network nodes 1 and 2, respectively. This result verifies that the learning-based algorithm can identify the under-attack node by deriving from the resource utilization efficiency. With the learned results, the learning-based algorithm allocates resources to the high efficiency nodes to obtain higher performance. As a result, the performance of the network slices will not be degraded significantly. This result verifies that the learning-assisted algorithm can mitigate the impact of the DoS attack by controlling the resource allocation. On the other hand, the baseline algorithm is unable to adjust the resource allocation under the DoS attack as shown in Fig.~\ref{fig:sys_allocation_detail} (b).


\section{Conclusion and Future Work }
\label{sec:conclusion}
In this article, we have discussed the needs and challenges of supporting cyber-physical systems with virtual network slices. By providing network slices with functional and performance isolation to various vertical services, the attack on a single slice may not affect the performance of others. The desired virtualization techniques should be capable to isolate the effect of attacks on the virtual resource layer without affecting the physical infrastructures. Besides, we have identified the security vulnerability of network slicing caused by the multi-domain resource virtualization.
Given the numerous attack types, e.g., DoS and man-in-the-middle, and the complicated influence on cyber-physical systems, e.g., performance degradation, intelligent solutions for identifying attacks, isolating attacks influence, and excluding attacks from the network are highly desired.
Since machine learning (ML) techniques have been successfully applied in various areas such as computer vision and robot control, utilizing emerging ML and developing learning based algorithms is promising to tackle various attacks on cyber-physical systems.
To address the security issue, we have presented the learning-assisted network slicing solution and analyzed the performance of the network slices under the denial-of-service (DoS) attack. The simulation results show that the learning-assisted network slicing solution is able to mitigate the impact of the DoS attack on the network slices. We have also presented the development of a small-scale testbed for evaluating network slicing solutions for cyber-physical systems.


%





\ifCLASSOPTIONcaptionsoff
  \newpage
\fi



\bibliographystyle{IEEEtran}

\begin{IEEEbiography}[{\includegraphics[width=1in,height=1.25in,clip,keepaspectratio]{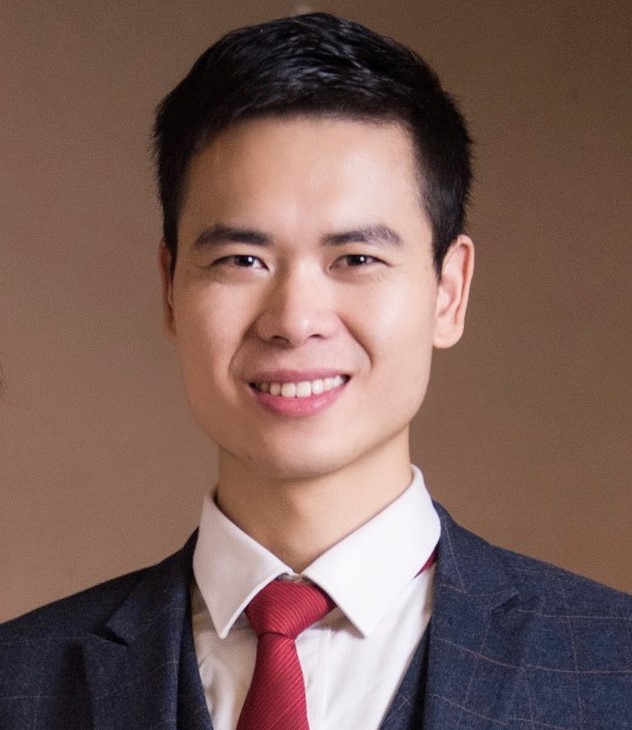}}]{Qiang Liu} received the M.Eng. degree from the University of Electronic Science and Technology of China in 2016. He is currently pursuing the Ph.D. degree in Electrical Engineering at the University of North Carolina at Charlotte. His research topic is machine learning for wireless edge computing networks.
\end{IEEEbiography}
\begin{IEEEbiography}[{\includegraphics[width=1in,height=1.25in,clip,keepaspectratio]{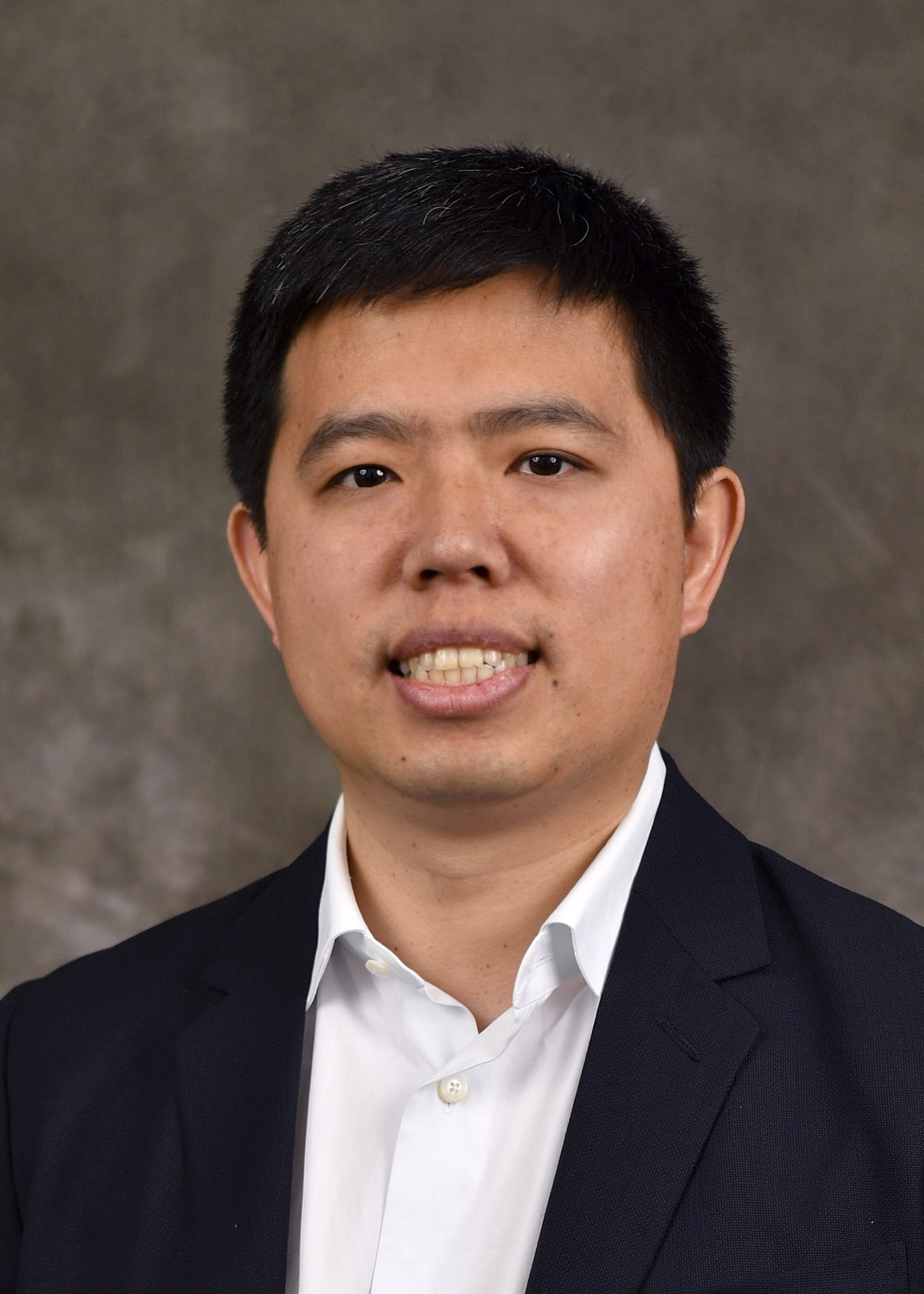}}]{Tao Han} (S'08-M'15)
received his Ph.D. in Electrical Engineering from New Jersey Institute of Technology (NJIT), Newark, NJ, USA. He is currently an Assistant Professor in the Department of Electrical and Computer Engineering at the University of North Carolina at Charlotte, Charlotte, NC, USA. He serves as an Associate Editor of IEEE Communications Letters. His research interest includes mobile edge networking, mobile X reality, 5G, Internet of Things, and smart grid.
\end{IEEEbiography}

\begin{IEEEbiography}[{\includegraphics[width=1in,height=1.25in,clip,keepaspectratio]{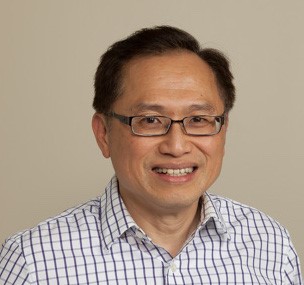}}]{Nirwan Ansari} (S'78-M'83-SM'94-F'09) is Distinguished Professor of Electrical and Computer Engineering at the New Jersey Institute of Technology (NJIT). He authored Green Mobile Networks: A Networking Perspective (Wiley-IEEE, 2017) with T. Han, and co-authored two other books. He has also (co-)authored more than 600 technical publications, over 280 published in widely cited journals/magazines. He has guest-edited a number of special issues covering various emerging topics in communications and networking. He has served on the editorial/advisory board of over ten journals including as the Associate Editor-in-Chief of IEEE Wireless Communications. His current research focuses on green communications and networking, cloud computing, drone-assisted networking, and various aspects of broadband networks.
He was elected to serve in the IEEE Communications Society (ComSoc) Board of Governors as a member-at-large, has chaired some ComSoc technical and steering committees, has been serving in many committees such as the IEEE Fellow Committee, and has been actively organizing numerous IEEE International Conferences/Symposia/Workshops. He has frequently been delivering keynote addresses, distinguished lectures, tutorials, and invited talks. Some of his recognitions include several Excellence in Teaching Awards, a few best paper awards, the NCE Excellence in Research Award, several ComSoc TC technical recognition awards, the NJ Inventors Hall of Fame Inventor of the Year Award, the Thomas Alva Edison Patent Award, Purdue University Outstanding Electrical and Computer Engineering Award, NCE 100 Medal, and designation as a COMSOC Distinguished Lecturer. He has also been granted more than 40 U.S. patents.
He received a Ph.D. from Purdue University, an MSEE from the University of Michigan, and a BSEE (summa cum laude with a perfect GPA) from NJIT.
\end{IEEEbiography}







\end{document}